%% file: main.tex
\documentclass[lettersize,journal]{IEEEtran}
\usepackage{amsmath,amsfonts}
\usepackage{algorithmic}
\usepackage{algorithm}
\usepackage{array}
\usepackage{textcomp}
\usepackage{stfloats}
\usepackage{url}
\usepackage{verbatim}
\usepackage{graphicx}
\usepackage{cite}
\usepackage{cleveref}

\usepackage{wrapfig}

\usepackage[framemethod=tikz]{mdframed}

\makeatletter
\let\MYcaption\@makecaption
\makeatother
\makeatletter
\let\@makecaption\MYcaption
\makeatother

\definecolor{Blue}{cmyk}{0, 0.87, 0.68, 0.32}
\definecolor{OliveGreen}{cmyk}{1, 0.50, 0, 0}
\definecolor{Red}{RGB}{219, 48, 122}
\definecolor{Orchid}{RGB}{218, 112, 214}
\definecolor{Teal}{RGB}{0, 128, 128}

\begin{document}

\title{Future Internet Congestion Control:\\The Diminishing Feedback Problem}


\author{\IEEEauthorblockN{Michael Welzl\IEEEauthorrefmark{1}, Peyman Teymoori\IEEEauthorrefmark{1}, Safiqul Islam\IEEEauthorrefmark{2}, David Hutchison\IEEEauthorrefmark{3}, Stein Gjessing\IEEEauthorrefmark{1} }
	\\\IEEEauthorblockA{\IEEEauthorrefmark{1}
		University of Oslo, Norway
	}
	\IEEEauthorblockA{\IEEEauthorrefmark{2}University of South-Eastern Norway
	}
	\IEEEauthorblockA{\IEEEauthorrefmark{3}Lancaster University, UK
	}
	
	}

\markboth{}%
{ }


\maketitle

\begin{abstract}
It is increasingly difficult for Internet congestion control mechanisms to obtain the feedback that they need. This lack of feedback can have severe performance implications, and it is bound to become worse. In the long run, the problem may only be fixable by fundamentally changing the way congestion control is done in the Internet. We substantiate this claim by looking at the evolution of the Internet's infrastructure over the past thirty years, and by examining the most common behavior of Internet traffic. Considering the goals that congestion control mechanisms are intended to address, and taking into account contextual developments in the Internet ecosystem, we arrive at conclusions and recommendations about possible future congestion control design directions. In particular, we argue that congestion control mechanisms should move away from their strict ``end-to-end'' adherence. 
This change would benefit from avoiding a ``one size fits all circumstances" approach, and moving towards a more selective set of mechanisms that will result in a better performing Internet. We will also discuss how this future vision differs from today's use of Performance Enhancing Proxies (PEPs).
\end{abstract}


\input{intro}
\input{cc-history}

\input{evolution}
\input{discussion}
\input{conclusion}



%
\bibliographystyle{IEEEtran}
\bibliography{literature}

\section{Biography Section}
%
%
%
%

\vspace{-1cm}

\begin{IEEEbiography}[{\includegraphics[width=1.05in,height=1.5in,clip,keepaspectratio]{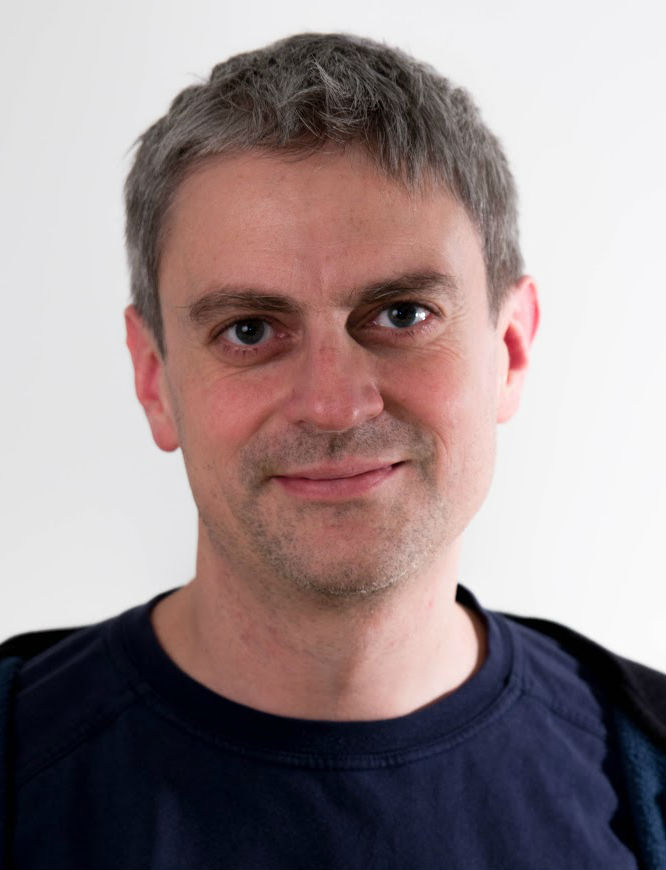}}]{Michael Welzl}
is a full professor at the University of Oslo,
Norway, since 2009. He received his Ph.D. and his habilitation
from the University of Darmstadt / Germany in 2002 and 2007,
respectively. Michael's main research focus is the
transport layer; he is active in the IRTF and the IETF.
\end{IEEEbiography}
\vspace{-1cm}

\begin{IEEEbiography}[{\includegraphics[width=1.05in,height=1.5in,clip,keepaspectratio]{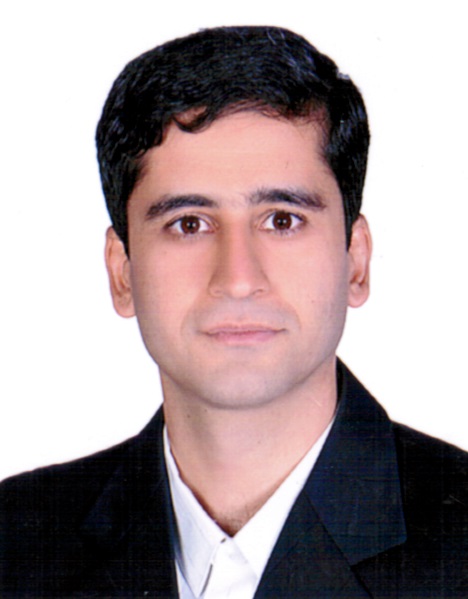}}]{Peyman Teymoori}
received his Ph.D. degree in computer engineering from University of Tehran, in 2013. Now, he is a researcher fellow in the Network and Distributed Systems group, Department of Informatics, University of Oslo, Norway. His research interests include the design and performance evaluation of computer network protocols, wireless networks, and recursive network architectures.
\end{IEEEbiography}
\vspace{-1cm}

\begin{IEEEbiography}[{\includegraphics[width=1.05in,height=1.5in,clip,keepaspectratio]{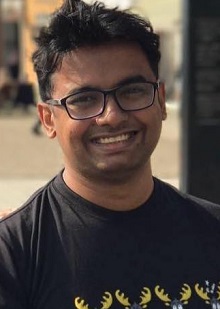}}]{Safiqul Islam} is an Associate Professor at the University of South-Eastern Norway. He
received his Ph.D. in Computer Science from the University of Oslo, Norway. His research interests include performance analysis, evaluation, and optimization of transport layer protocols. He is active in the IETF and IRTF where he has contributed to several IETF/IRTF Working Groups.
\end{IEEEbiography}
\vspace{-1cm}

\begin{IEEEbiography}[{\includegraphics[width=1.05in,height=1.5in,clip,keepaspectratio]{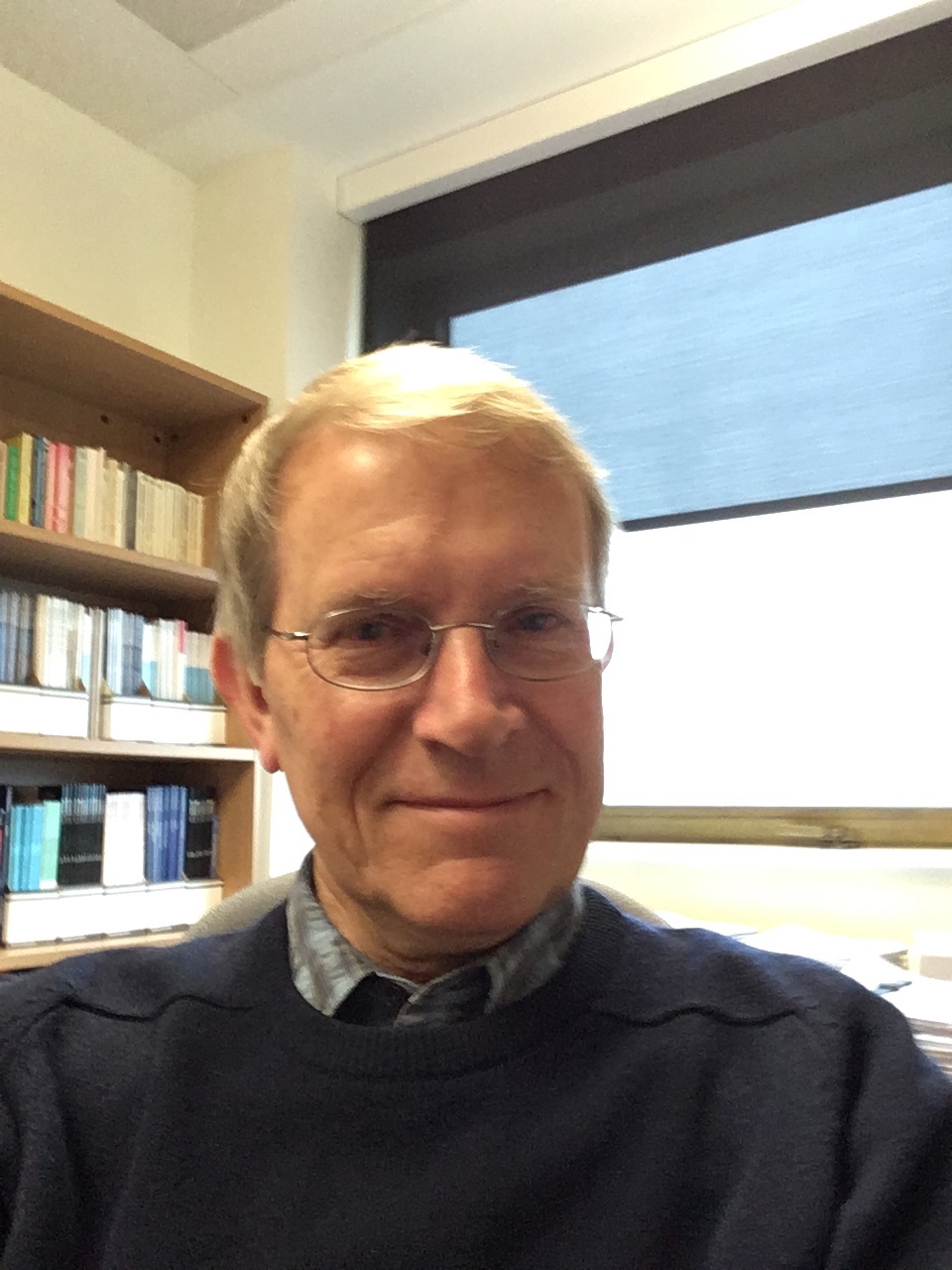}}]{David Hutchison}
is a Distinguished Professor of Computing at Lancaster University, UK, and the Founding Director of InfoLab21. His work is well known internationally in a range of areas including Quality of Service for computer networks, programmable networking, multimedia and content distribution networks, and testbed developments. His most recent research focuses on the resilience of networked computer systems, and the protection of critical infrastructures and services.
\end{IEEEbiography}
\vspace{-1cm}
\newpage
\begin{IEEEbiography}[{\includegraphics[width=1.05in,height=1.5in,clip,keepaspectratio]{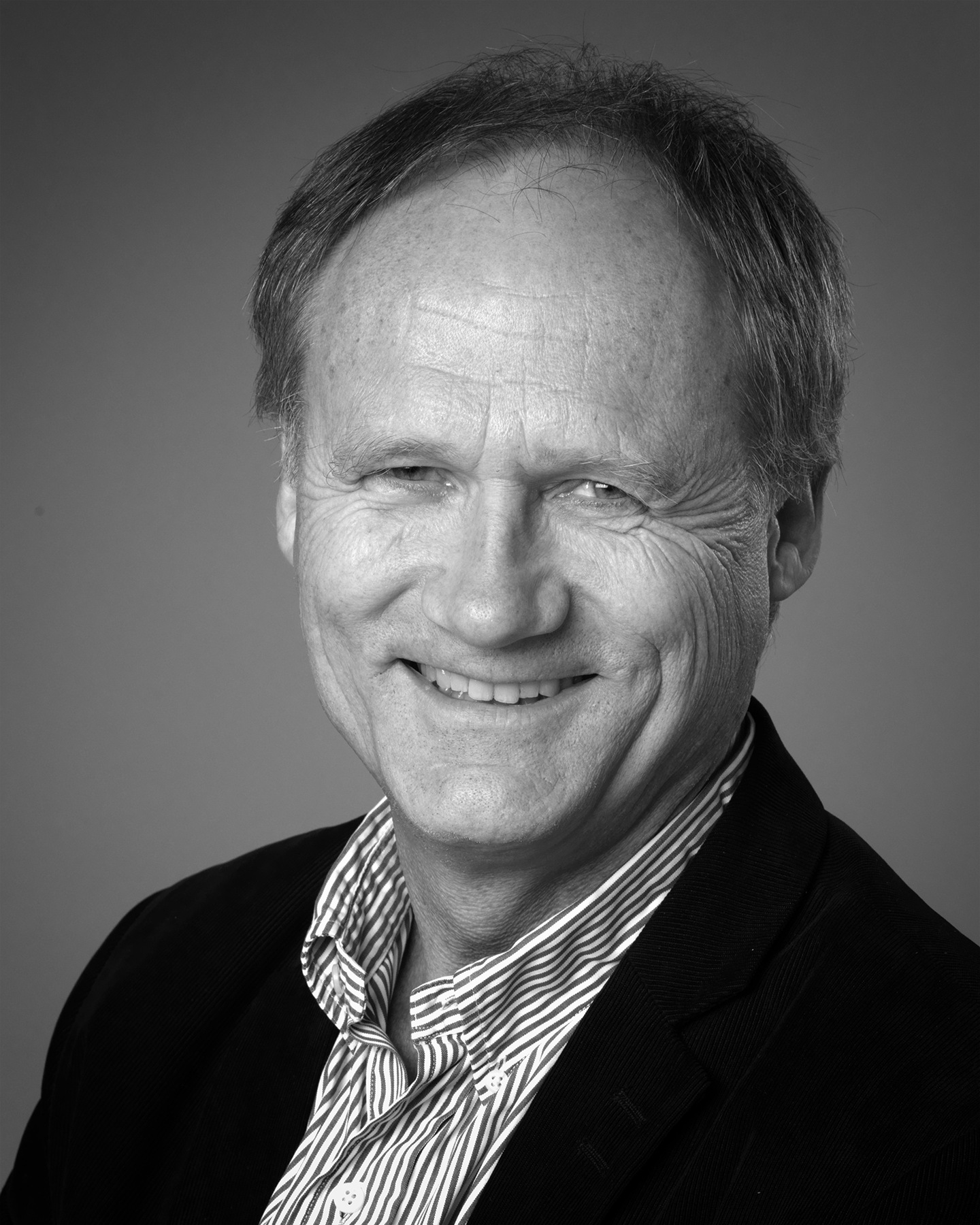}}]{Stein Gjessing}
is an emeritus professor of Computer Science in Department of Informatics, University of Oslo. He received his  Dr. Philos. degree in 1985 form the University of  Oslo.  He has worked with  object-oriented concurrent programming, computer  interconnects  and  computer  architecture  for  cache  coherent shared  memory,  with  DRAM  organization,  with  ring  based  LANs  (IEEE Standard  802.17)  and  with   IP  fast reroute.
\end{IEEEbiography}

\end{document}

%% file: intro.tex
\section{Introduction}

\IEEEPARstart{C}{ongestion} control is a network function that aims at operating the network at a state where all sources send with their ideal sending rates. Bottleneck capacities should be fully utilized, 
queues should mostly be empty, and rates should be allocated to senders according to some definition of ``fairness''.

The term ``congestion control'' was coined at a time when experimental networks occasionally were overloaded, and solutions were needed that would prevent such overload from happening. The term indicates the necessity of a control loop, which naturally entails that the sending rates of traffic sources can increase as well as decrease---but arguably, ``congestion control'' also sounds as if there has to be a congestion problem to solve in the first place. As we will discuss in the next section, this has not typically been the case in the Internet during the last two decades: capacities have grown, networks tend to be overprovisioned, and the prevailing problem is \emph{underload} rather than \emph{overload}.
In hindsight, ``rate control'' might have been a more appropriate term.

Compared to overload, underload is harder to notice, and less likely to be perceived as a problem.
Indeed, limited utilization may often be good for network operators, as they need some leeway to cope with sudden traffic surges from major events with an unusually large number of participants (e.g. the football World Cup). So, why is underload a problem at all?

Consider web traffic. It consists of typically short, congestion controlled, reliable data transfers, and it is latency-critical; the delay observed by users depends on the completion time of transfers. This delay affects a multitude of web-based applications---not just access of static pages, but also services such as online road maps, various cloud applications, and e-commerce, for example. These short transfers can experience unnecessary delay even when there is plenty of available capacity, and we need to reduce such delays by better utilizing the network capacity. This is a case where time is money: users of websites are quick to turn away when they experience latency. Several companies have attempted to calculate the related revenue loss. For example, already in 2006, Amazon estimated that every 100~ms of delay costs 1\% of sales~\cite{amazondelay}.

The root of the problem is that the original Internet congestion control mechanism was designed to operate in circumstances where traffic loads would normally be moderate or large compared to link capacities. When networks are significantly underloaded, the mechanism delays packets unnecessarily because it \emph{lacks feedback}: unaware of the excessive network capacity, it continues to apply the regular algorithm.

The situation will be explained further in the next section, where we will discuss how
Internet congestion control has changed over the last decades. As we will see, missing feedback became an issue at an earlier stage of Internet development, and it was solved---yet, this prior solution does not solve today's problem. \Cref{sec:evolution} examines factors that make the present situation problematic: because of how the network infrastructure is evolving, the problem is bound to become worse, and fixing it may require a radical change of the way Internet congestion control operates.
We will therefore discuss appropriate updates to congestion control, and what implications such changes might bring, in \Cref{sec:discussion}.

%% file: cc-history.tex
\section{What problem does congestion control solve?}
\label{sec:history}

Responding to the prevalent problems at the time, the primary focus of Internet congestion control has changed multiple times over the course of three decades, which allows us divide the related research into three major phases. We
discuss congestion control in the context of its traditional vehicle: the ``Transmission Control Protocol (TCP)''. However,
our observations are general, and therefore also apply to other protocols such as ``Quick UDP Internet Connection (QUIC)''~\cite{rfc9000}.

%
\subsection{Phase 1 (the 1990's):\\ Keeping the Internet Operational}

Congestion control was added to TCP in response to a global Internet ``congestion collapse'' in the late 1980's; the introduced behavior added a ``congestion window (cwnd)'' as an extra limitation on the number of packets to send per round-trip time (RTT).
From an initial value called ``Initial Window (IW)'', TCP gradually increases its cwnd until it notices a sign of congestion, as a result of a growing queue at the bottleneck router---at the time, losing a packet when the queue reaches its maximum size was the only indicator. Then, TCP backs off, and the cycle repeats. The initial increase is exponential,
dubbed ``slow start'', and later increases, dubbed ``congestion avoidance'', are linear. Very severe
congestion can put TCP back to its initial state, where it begins with slow start again.
Because TCP congestion control shifted the whole Internet from a failure mode back into being ``normally'' operational, in the following years, the primary focus was on preserving this relatively stable operation.

\begin{figure}[ht]{}
    \centering
    \includegraphics[width=0.95\linewidth]{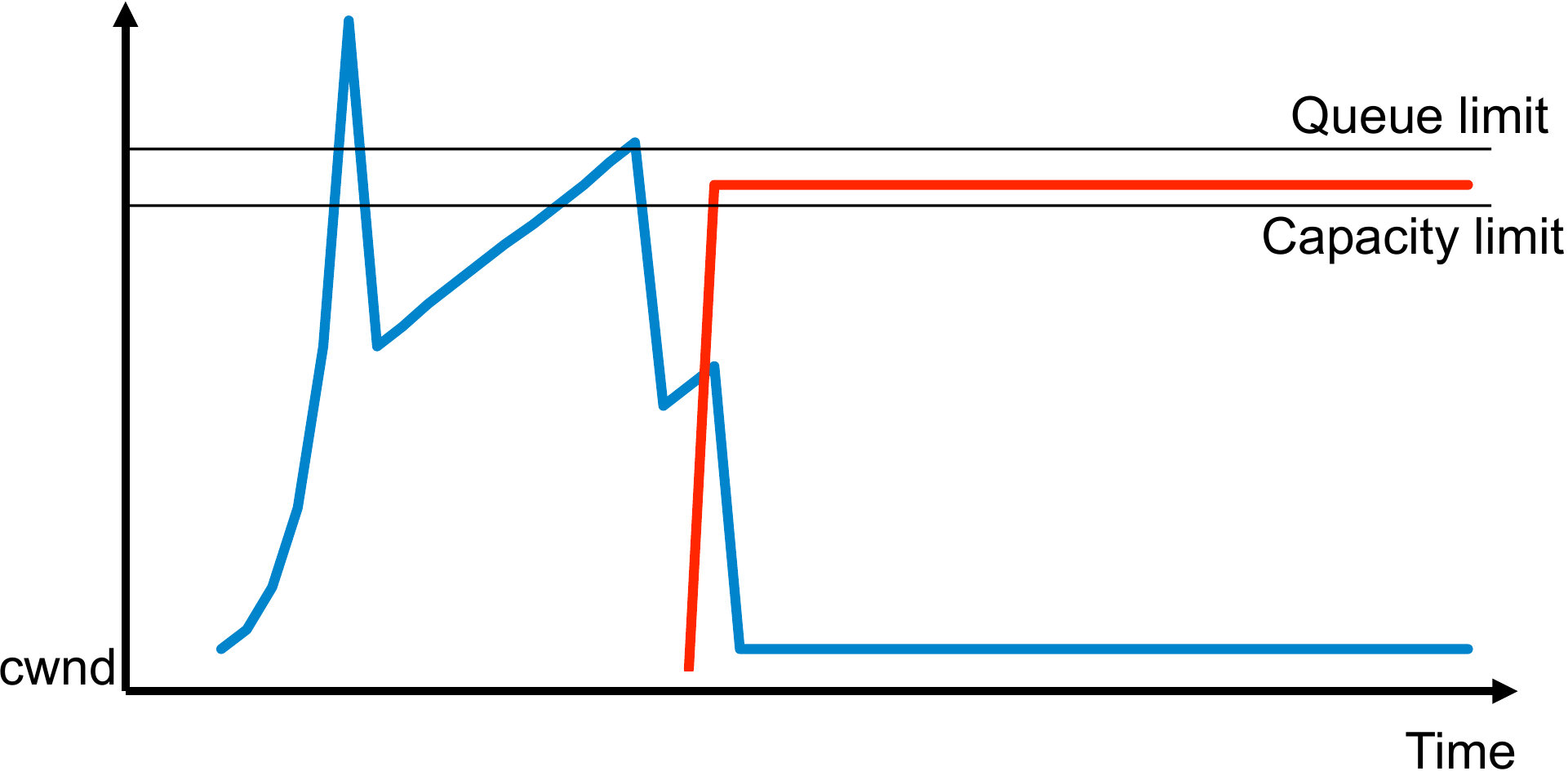}
    \caption{1990's concern: an aggressive unresponsive flow (red line) starves a TCP connection (blue line).}
    \label{fig:problem-1990s}
  \end{figure}

In the 1990's, work on rate-adaptive multimedia applications flourished, and there was a wish to offer such applications a suitable behavior in terms of congestion control. 
These applications needed some form of compatibility with TCP, as an unresponsive application using the ``User Datagram Protocol (UDP)'' can easily render a TCP connection (or even a large number of them) unusable. This concern is depicted in Figure\,\ref{fig:problem-1990s}.
  
To handle this problem, the notion of ``TCP-friendliness'' was established. It became acceptable to propose different congestion control methods which yield fewer quality fluctuations and less jitter than a TCP-like behavior, as long as these congestion control mechanism would, on average, not exceed the rate of a TCP connection under similar circumstances.

\subsection{Phase 2 (the 2000's):\\ Working With High-Speed Links}

\begin{figure}[ht]{}
    \centering
    \includegraphics[width=0.95\linewidth]{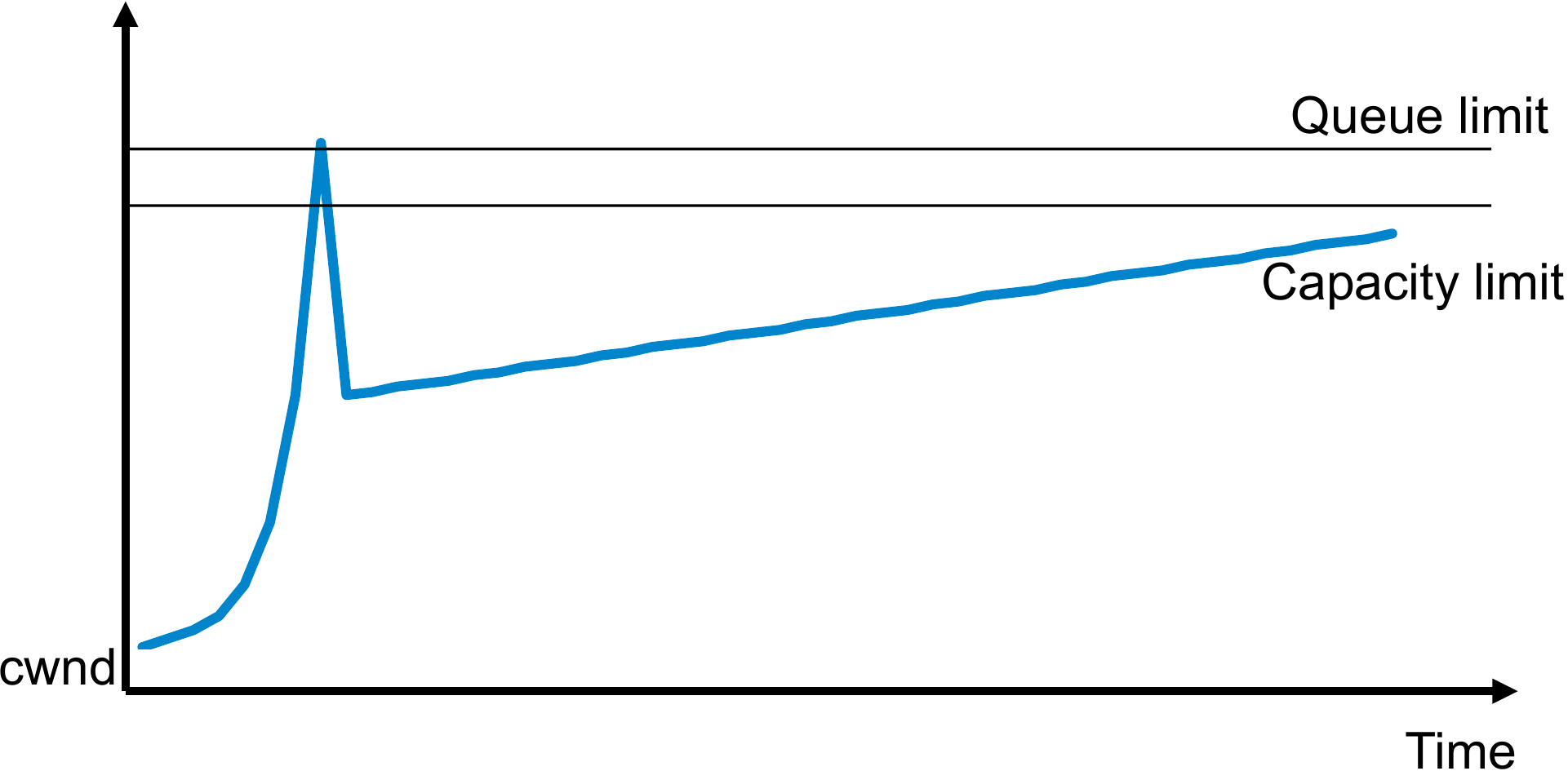}
    \caption{2000's concern: the bandwidth$\times$delay product is large (diagrams are not to scale), and TCP is too slow to reach the capacity limit.}
    \label{fig:problem-2000s}
  \end{figure}

The early 2000's saw an increase of network capacities up to a Gigabit per second, combined with a desire
to use the Internet for large-scale distributed (``Grid'') computing. For such uses, it
was necessary to ship very large amounts of data. 
TCP became a bottleneck: the additive increase of TCP's cwnd does not scale with
capacity, which means that it may sometimes
take a standard TCP flow up to an hour or more to fully saturate the bottleneck when
the Bandwidth\,$\times$\,Delay Product (BDP) is large.
Even large data transfers would often not
last long enough for TCP to saturate the capacity and obtain congestion feedback once under such circumstances.
As Figure~\ref{fig:problem-2000s} shows, the network was underloaded. 
  
Increasing the sending rate more quickly is not ``TCP-friendly'', however, and
thus, mechanisms for large BDP networks needed an 
``emergency brake'', to ensure that they only are
more aggressive than standard TCP when the packet loss ratio is low. 
%
One such mechanism, ``Binary Increase Congestion control TCP (BIC-TCP)'', 
%
was made the default in Linux in mid-2004.
The Internet did not melt, but the new Linux behavior was significantly more aggressive
than the behavior of other operating systems.
This situation was soon improved when 
the slightly less aggressive CUBIC mechanism 
was chosen as the new default.
Like other competitors of the time, neither BIC nor CUBIC is ``perfect'',
but it appears that CUBIC performed well enough to make the ``large BDP problem'' less problematic 
for the following years---or, perhaps, it was also due to the changing Internet usage.

\subsection{Phase 3 (since the 2010's): Minimizing Latency}
\label{sec:history-phase3}

\begin{figure}[ht]{}
    \centering
    \includegraphics[width=0.95\linewidth]{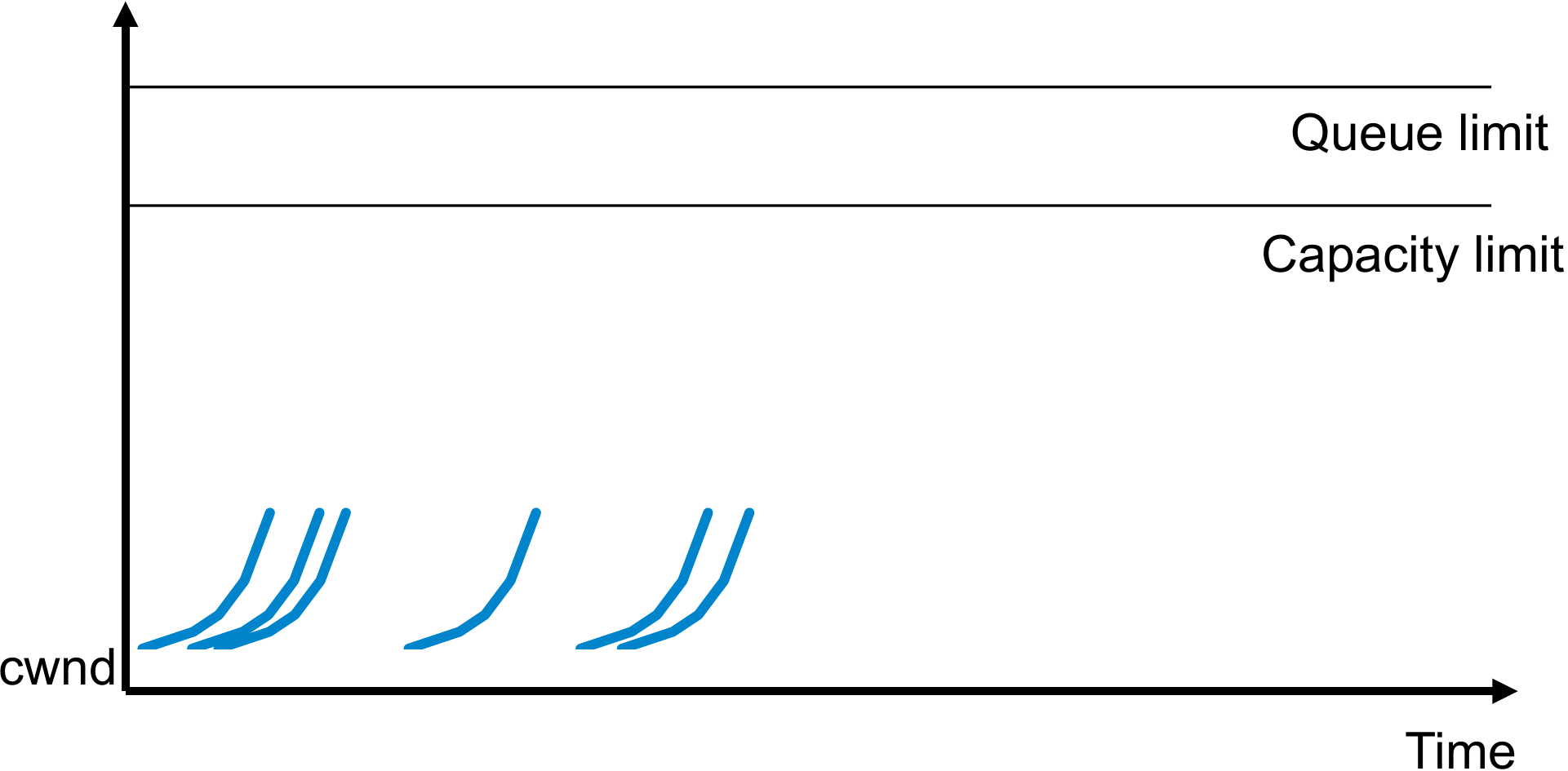}
    \caption{2010's (and later) concern: flows cannot probe for the capacity limit because they are too short. Reducing latency is the primary goal.}
    \label{fig:problem-2010s}
  \end{figure}

Today, the Internet is a cornerstone of modern society. We use it to work from home, access social media,
chat with friends and family, play games and watch movies or TV. From the network's point of view,
much of this communication is short-lived. 
Web pages are transmitted
as many short connections, often from a variety of servers. Messengers often only transfer
text or images which easily fit in a handful of packets.
As Figure~\ref{fig:problem-2010s} shows,
these short or broken transfers often do not last long enough to probe for the available capacity
even once (we will confirm this with measurements in the next section), yet, for most web based
applications, they should be finished as quickly as possible. No current congestion control
mechanism lets a sender increase its cwnd faster at this initial stage.

%


%
%
This means that the optimization goal is: keep the on-path latency low (by avoiding queuing delay)
while \emph{quickly} making use of the available capacity. We will discuss how this could be
done in \Cref{sec:discussion}---but first, we confirm the importance and timeliness of solving this
optimization problem by taking a brief look at the development of the Internet
infrastructure as well as the typical properties of Internet traffic. 

%
%

%% file: evolution.tex
\section{Why is there a feedback problem?}
\label{sec:evolution}

We will now examine two factors that make the the present situation particularly worrisome. First, the way
the infrastructure has been evolving gives TCP an increasingly large operational space in which it does not see
any feedback at all. Second, most TCP connections are extremely short. As a result, it is quite
rare for a TCP connection to even see a single congestion notification during its lifetime.

\subsection{The Evolution of Internet Connection Speeds}

\begin{figure}[t]
\centering
\includegraphics[width=1.0\columnwidth]{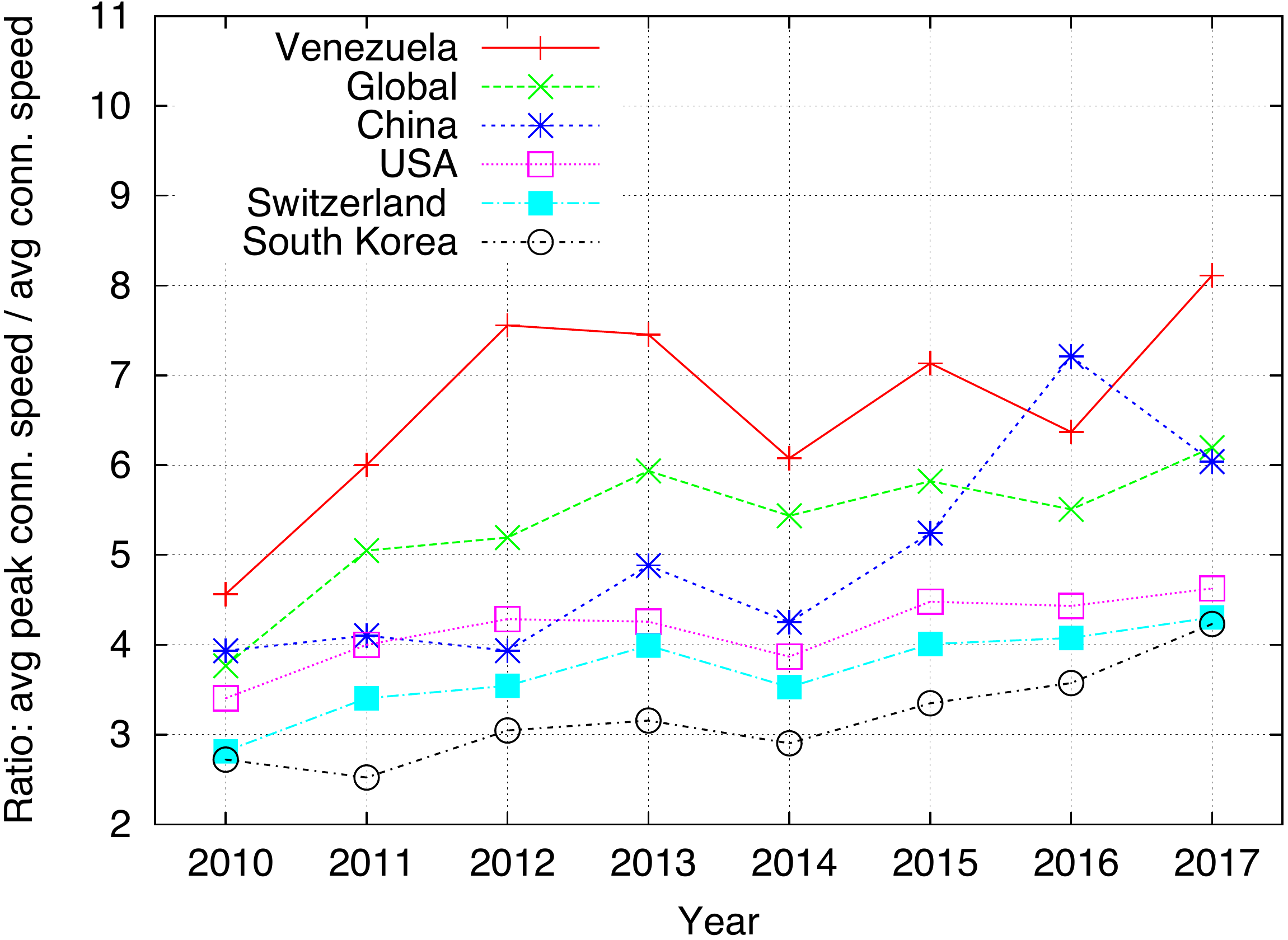}
\caption{Internet connection speed developments and trends according to Akamai's ``State of the Internet'' reports.}
\label{fig:akamai}
\end{figure}

Figure\,\ref{fig:akamai} shows the development of the ``average peak'' vs. average connection speeds according to Akamai's ``State of the Internet'' reports in the time frame 2010-2017 (newer reports do not contain this information). The ``average peak'' represents an average of the maximum measured connection speeds. From the countries where data was available for the full 2010-2017 time frame, we chose 
South Korea and Switzerland as high-capacity examples, and China and Venezuela as lower-capacity cases, respectively, in addition to the USA and the global average.
Altogether, an upward trend of the peak vs. average ratio is noticeable. In fact (not evident from Figure\,\ref{fig:akamai} but available in the reports), from 2010 to 2017, the average capacity has roughly quadrupled, both globally and in the USA---but the global average peak capacity in 2017 is 7 times the value of 2010 (5.4 in case of the USA).
This shows that the Internet not only becomes faster: the range between low-end and high-end connectivity grows. The reasons are not documented, but this may be the effect of Internet connections being upgraded at different rates---i.e., some links are upgraded, but some legacy links with lower capacities are still in place.

\begin{figure}[t]
	\centering
	\includegraphics[width=1.0\columnwidth]{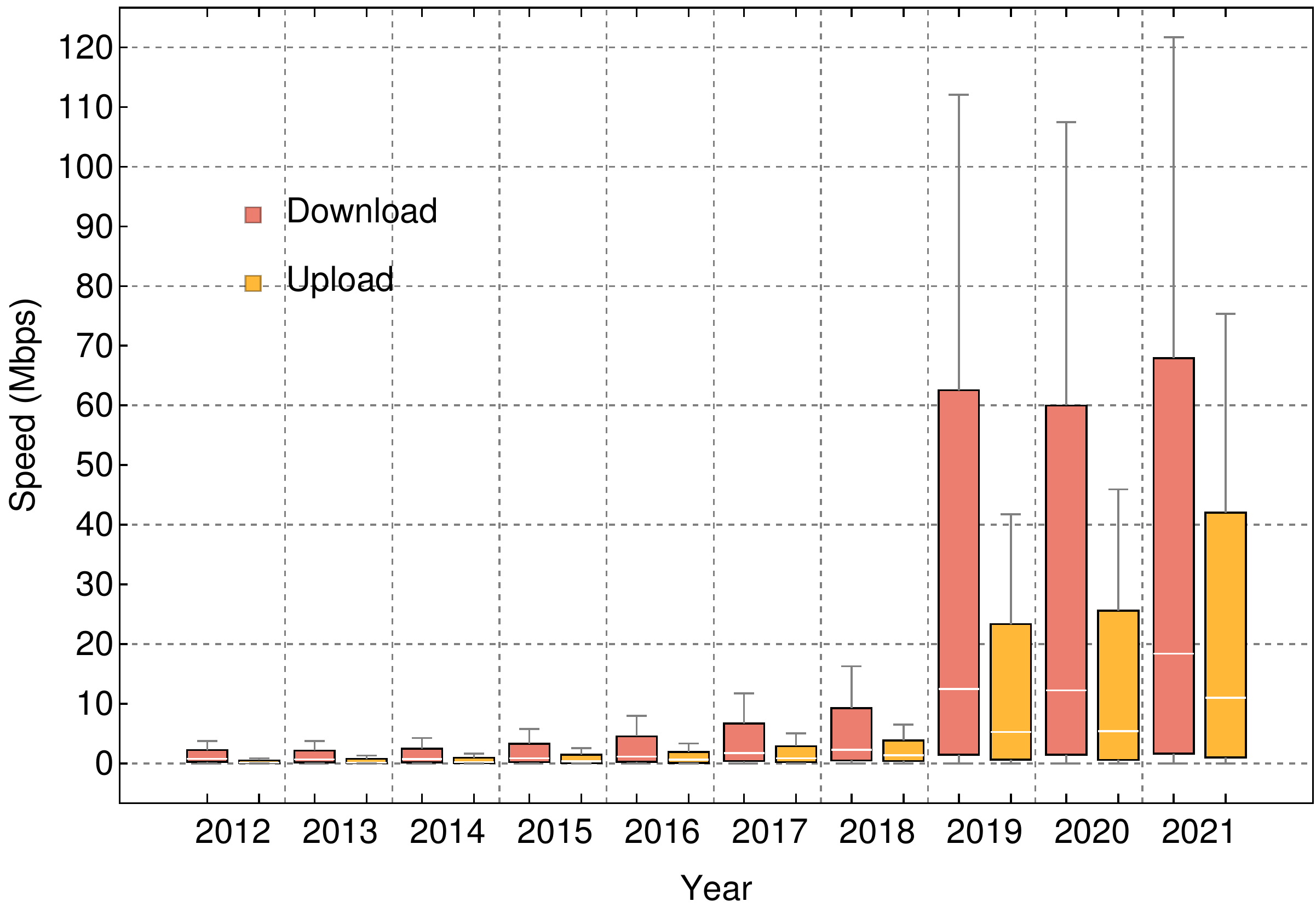}
	\caption{Internet throughput 
	ranges and developments for 90 countries according to Measurement Lab data sets. The sudden increase in 2019 is due
	to the measurement campaign switching to CUBIC congestion control, which eliminates the problem shown in Figure\,\ref{fig:problem-2000s}.}
	\label{fig:mlab}
\end{figure}

%
%
%
Measurements confirm this: Figure~\ref{fig:mlab} illustrates the evolution of end-to-end Internet throughput ranges 
according to Measurement Lab data sets. 
The data were collected from 2012 to November 2021 in over 90 countries by a tool called Neubot~\cite{de2008neubot} that ran daily in the background of up to 4500 static volunteer user computers and periodically tested the network performance. Some values
are very small because many of the countries in the study have poor Internet connectivity.
The lower and upper ends of the boxes in the figure represent the 25\% and 75\% percentiles, respectively. Outliers are not presented for the sake of clarity. Referring to the figure, we observe that the range of link capacities is widening. For example, 
the difference between 25\% and 75\% percentiles from 2012 to 2021 is 0.328, 0.499, 0.621, 0.988, 1.29, 1.91, 2.38, 16.2, 18.1, 29.6\,Mbps, for upload, and 1.37, 1.4, 1.6, 2.2, 3.1, 4.5, 6.2, 44, 42.2, 48.1\,Mbps, for download, respectively. From 2019, when CUBIC was enabled for these measurements in the Neubot tool, the upper end of the measured values increases drastically, but the lower end is only marginally affected. 

%
%

\subsection{Wireless connectivity}

As we increasingly use wireless links to connect to the Internet, the development of wireless
technology has important ramifications for TCP/IP protocols too. One issue stands out in particular:
while it was once correct to assume that the capacity of a bottleneck along an end-to-end
path is relatively stable, this is no longer the case.

For example, very high data rates can be achieved using millimeter Waves (mmWave), which are adopted by new technologies such as 5G, IEEE 802.11ay and vehicular communications as the demand for high-speed mobile Internet access is growing---but mmWave communication can be obstructed by obstacles, causing sudden outages and high fluctuations in physical data rate. Tests that use capacity fluctuations from real-life mmWave traces show that this is very problematic for state-of-the-art congestion control mechanisms~\cite{cc-mwave-measurement}: a capacity reduction can cause cwnd to become very small, yet the end-to-end control loop is not informed when the capacity is increased again. This can lead to sustained underload. Also, the delay at which an adjacent queue is drained changes with the available capacity, affecting delay measurements.

\subsection{Properties of Internet traffic}
\label{sec:trafficproperties}

Internet traffic has long been known to consist of many short flows (``mice'': web browsing etc.) which compete with only a few very long flows (``elephants'': long file downloads etc.).
A recently published measurement study finds that, in a large dataset
consisting of traffic traces that were captured on a Tier-1 Internet Service Provider (ISP) backbone in Chicago between
June 2008 and March 2016, 85\% of all TCP connections carry between 100\,B and 10\,kB of data~\cite{internetflow}.
In a measurement of one week in 2017 in the mobile search service of Baidu, Inc., 
80.27\% of all TCP connections terminate in slow start~\cite{learning-iw}.


To confirm these findings, and to investigate how often short flows
encounter packet loss, we obtained data from the Japanese ``WIDE'' project's MAWI Working Group archive.
This archive contains daily traffic traces 
captured on
a 1\,Gbps transit link of WIDE to an upstream ISP.
\Cref{tab:losstable} summarizes our analysis of packet losses in these traces, which we identify in two ways, since losses can happen before or after the measurement point. To cover the first (before) case, we conclude that a flow experienced a loss when we see a retransmitted data packet. For the second case (losses after the measurement point), we consider three duplicate acknowledgments as an additional indication of loss. In this way, there is only one exceptional case of losses in end-to-end TCP connections that we miss: ``tail losses'' (loss of the last data packet in a connection) that happened before the measurement point. (Since no further 
acknowledgments would arrive, these retransmissions are caused by a timeout.)

In this example of real Internet TCP traffic, 85.5\% of all flows have no re-transmissions, i.e. they experienced no packet loss.
Because marking a bit in packets instead of dropping them (a scheme called
``Explicit Congestion Notification (ECN)'') is negligibly rare,
in all of these cases, the congestion control
mechanism never even managed to probe for the available capacity before the transfer ended.
It is important for the remaining large flows to carry out appropriate congestion control, or results could
be devastating for the many short flows that might see their packets enqueued or even dropped.

\begin{table}[t]
    \caption{Measurements---15-minute long traces
    from each month from January 2019 to November 2021 at the transit link of the Japanese WIDE backbone network.}
    \centering
    \begin{tabular}{lc}
       Total number of flows  &  2555005 \\ 
        Flows with re-transmissions (client to server) &  317754 (12.44\%) \\  
        Flows with re-transmissions (server to client) &  52738 (2.06\%) \\  
        Total flows with re-transmissions & 370492 (14.50\%) \\ 

    \end{tabular}   

    \label{tab:losstable}
\end{table}

We have already explained that the inability of TCP to saturate the network's capacity
incurs delay for web applications. However, this
shortcoming of congestion control is not equally important for all applications. Netflix,
for example, transmits data in short bursts, and it is common to see pauses of multiple seconds between them~\cite{adhikari-varvello-2012}.
As long as data from these bursts reach the receiver in time to keep the playback buffer from draining,
there is no problem. Since Netflix users can already watch videos at
the highest quality level that is being offered for their output devices, a better
congestion control mechanism would not make any difference.

The following example highlights the significance for web traffic. We accessed
the front page of \texttt{cnn.com} with Google Chrome (with QUIC disabled) and determined
the length of the longest TCP data transfer from the main server (which plays a role for the
overall delay in presenting the page). This transfer consisted of 948 data packets in the server-client direction and showed no noticeable packet loss.
Due to the initial ``Transport Layer Security (TLS)'' handshake, it is not easy to see the actual IW that the \texttt{cnn.com} server
used in our test. With the common choice of IW=10, it would take
at least seven round trips to finish this transfer. At the beginning of the 6th round, cwnd would
be 320, and by the end of this round, 630 packets would have
been transmitted. The value of 320 certainly worked in our test: within 948 packets, even IW=1 would have
allowed cwnd to grow beyond 320, with no packet loss incurred. This value could, in principle, have been used
from the start, allowing the transfer to terminate within three round-trips, i.e. in less than half of the previous time.
(Note that using such a large cwnd would probably require a regime to ``pace'' traffic, i.e. interject pauses in between packets, because large traffic bursts can cause a router queue to overflow, causing packet loss.)

%% file: discussion.tex
\section{The way forward}
\label{sec:discussion}

What is the road ahead for Internet congestion control in the context of its diminishing feedback problem?

The large majority of previous congestion control research has focused on improving the
congestion avoidance phase of TCP, which begins upon the first packet loss. Moreover,
it has been common to assume a ``greedy'' sender, i.e. an application which always
sends as much data as allowed by the congestion control mechanism.
Given how rare packet losses and long-term ``greedy'' applications really are, this
kind of improvement may no longer be worth the effort, and the focus should probably
shift to improving the initial start-up behavior.

We will now take a closer look at three directions for such improvements: changing the end-to-end
control, letting devices in the network help, and changing the path itself.

\subsection{End-to-end congestion control}

To some degree, the shift of focus has already happened.
Since underload is primarily disadvantageous for web traffic, it is not surprising
that Google has already developed a number of fixes for this problem, with the
goal to reduce the completion time of short Internet
data transfers. Several of these fixes are incremental changes to existing
standards, and have been standardized in the
Internet Engineering Task Force (IETF). But how far can these gradual improvements take us?

In the absence of packet loss, the completion time of TCP transfers primarily
depends on three factors: i) the number of management round-trips, ii) the RTT
and iii) TCP's slow start behavior.
Recent work related to TCP (``TCP Fast Open (TFO)'')~\cite{rfc7413},
TLS~1.3~\cite{rfc8446} and QUIC~\cite{rfc9000}) has brought the number
of management round-trips down, in some cases even to zero; we may
well have reached the limit of optimization for this aspect.
Regarding TCP's slow start behavior,  Google has successfully lobbied for an increase of the standard
IW from 4 to 10 TCP packets, and Internet servers now use a large range of different
values, with common choices as low as 1 and as large as 48,
or even 100
in some cases~\cite{iw-measurement-cdn}.

If the Internet would uniformly become faster everywhere, TCP could gradually
be improved by scaling the IW along over the years---however, as we have
discussed, this is not the case: it appears
that low-end capacity links do not disappear as quickly everywhere
as high-end capacity links become available in some parts of the world.
Since transport protocols like TCP and QUIC are worldwide standards, they
must be able to cope with the increasingly difficult situation of a growing operational
range in which there is no feedback.

This points at a need to make an informed choice instead. One possibility is to
try to learn from past success or failure, and to assume some correlation between
a newly starting data transfer and previous transfers. This has, for example, been
done by the authors of~\cite{learning-iw}, who report positive results obtained
with a reinforcement learning based IW adjustment strategy
during a year of production use for the mobile search service of Baidu, Inc.

The relatively new mechanism ``Bottleneck Bandwidth and Round-trip propagation time (BBR)'' offers potential for using the network capacity more suitably to avoid unnecessary latency due to its use of pacing~\cite{cardwell-iccrg-bbr-congestion-control-01}. 
Because large packet bursts are likely to cause queues to overflow, pacing is advisable when trying to speed up short transfers, whether this is done by changing the IW or by adapting the cwnd increase behavior. Currently however, BBR does not propose a change to the default IW value and also does not initially increase its rate any faster than the common slow start mechanism.

%

\subsection{Network-assisted congestion control}


Future congestion controls could leverage two essential methods
to cope with this increasingly difficult situation:

\begin{enumerate}

\item {\bf Explicit feedback:} The problem of diminishing feedback is intrinsically connected to
its \emph{implicit} nature---measurements of delay and packet loss, as a
result of a growing queue. Explicit feedback can be provided earlier, before 
a queue attains a significant length.

More than a decade ago, researchers proposed a handful of mechanisms that are based on
a joint sender-router control design with multi-bit explicit feedback (e.g., the ``eXplicit Control Protocol'' (XCP)~\cite{xcp}).
These mechanisms worked much better than the more common implicit feedback based
schemes---including a much faster startup behavior---but their reliance on support from all routers along the path made them unsuitable for practical use in the Internet. Even the much simpler ECN scheme, where routers set a bit to indicate congestion instead
of dropping a packet, has a long history of deployment failures. This highlights the difficulty to deploy explicit feedback
in support of an end-to-end control loop. However, explicit feedback schemes may work well on shorter path segments.


\item {\bf Faster feedback:} The reaction speed
of any congestion control mechanism is in the order of RTTs---so,
to react faster, the RTT needs to be shorter. This could be achieved by
putting the control close to the bottleneck.
Also, short-RTT loops could then carry data from multiple end-to-end flows. Such
aggregate control would avoid unnecessary
side-effects from competition between end-to-end flows, and it would allow to immediately
assign an ideal share of the current aggregate rate to a newly arriving short flow.
\end{enumerate}

Control loops are already shortened today
by Performance Enhancing Proxies (PEPs) which, in their simplest form, split a
TCP connection in two, claiming to be the receiver towards the sender and claiming
to be the sender towards the receiver~\cite{rfc3135}. With such a connection splitting device in place,
it is, in principle, not necessary for the same congestion control mechanism to be
uniformly deployed across the whole end-to-end path.
PEPs therefore allow better use of links that have a fluctuating capacity, such
as a mmWave link layer.

PEPs, however, come with a number of problems. For one, by ``cheating'' TCP, they need to conflate congestion control and reliability. As they reduce the length of the congestion control loop, they must also buffer and retransmit packets---such buffers can add delay. Because they take decisions based on packet headers that should not normally be inspected within the network, and seeing something unexpected in these headers can cause failures, PEPs also contribute to the ``ossification'' of the Internet architecture (the difficulty of changing network protocols). This is one of the reasons why QUIC encrypts not only the payload but also large parts of the packet header~\cite{rfc9000}.

The ``Multiplexed Application Substrate over QUIC Encryption (MASQUE)'' IETF working group
develops a set of standards in which end systems explicitly communicate with proxies.
Such proxies could, at least in theory, carry out certain functions that were previously
only common for TCP PEPs, but they would no longer be confined to the limitations
imposed by having to ``cheat''.
A similar approach already exists for TCP~\cite{rfc8803}; it enables proxy
support for Multipath TCP, in line with the Access Traffic Steering, Switching, and Splitting (ATSSS) service being specified within the 3rd Generation Partnership Project (3GPP).
However, so far, changing congestion control to improve
the startup behavior has not been a focus of this work.

%

\subsection{Changing the path}
\label{sec:mec}

The RTT can be minimized by placing
content as close as possible to its consumer using a ``Content Distribution Network (CDN)''.
Beyond serving static content, ``Multi-access Edge Computing (MEC)'' puts a variety of
applications closer to their users---to the ``edge'' of the network.
The shorter, more homogeneous MEC path may directly use explicit feedback
without the need for a proxy; this is also envisioned in the ETSI MEC standard~\cite{etsi}.

With CDNs and MEC, the underload problem of Internet congestion control does not disappear, but
its effect may become minuscule. For the \texttt{cnn.com} example that we presented in Section~\ref{sec:trafficproperties},
better congestion control may have reduced the number of round-trips from 7 to 3. If the RTT to \texttt{cnn.com}
is 100\,ms, this means a reduction by 400\,ms---but when the RTT to the server is only 10\,ms, better congestion
control could only reduce the delay by 40\,ms.

Multi-access (multi-path) congestion control can also play a role for latency reduction---for example,
the ``Siri'' application by Apple Inc., is known to duplicate messages across multiple paths and to use
the earliest arriving one in order to speed up processing. Here, again, the underload problem is relevant
for each of the paths in use, even though the latency reduction per path
will be relatively small.

%% file: conclusion.tex
\section{Conclusion}
\label{sec:conclusion}

The focus of this paper is whether Internet congestion control needs to change. Given the vastly increasing capacity of the Internet infrastructure and typical application traffic behavior with mostly short or interrupted flows, we argue for improved congestion feedback with a shift away from the current end-to-end approach; this will improve performance.  

The Internet ecosystem is also evolving towards a more fragmented set of networks, with differing commercial goals and imperatives. Our assumption, however, is that the basic architecture and protocols in each will remain, but we conclude that congestion control mechanisms in each network instance must have the ability to be varied---i.e. moving away from a ``one size fits all circumstances'' approach.



For some applications, 
latency benefits have been made with the introduction of QUIC, which uses UDP instead of TCP, thus incurring fewer protocol round trips~\cite{rfc9000}. 
We believe that the focus should next be on replacing QUIC's congestion control mechanism, which would be a more fruitful source of latency saving---especially in a large-capacity network when it is significantly underloaded.

Nevertheless, today, any congestion control mechanism in QUIC or TCP operates end-to-end, which we claim brings disadvantages. Further investigation of PEPs is needed---looking at turning them into  ``good network citizens'' and deploying a mechanism that reacts to explicit feedback on a per-path-segment basis. Today, PEPs are almost exclusively used with links that have a very peculiar behavior, e.g. a satellite or mmWave link. We believe there is a more general use for PEPs, e.g. to increase the chance of discovering a workable IW value over time by breaking paths into shorter segments.

We hope this paper will persuade our peers that it is worth considering these issues, to debate and experiment with alternative designs, and help prepare the Internet to shift towards a future in which unnecessary latency is substantially reduced or removed, and congestion control
is no longer routinely implemented end-to-end.